\documentstyle[preprint,aps,eqsecnum]{revtex}
\begin{document}
\preprint{UCD 93-19,  July 1993}
%\draft
\title{Adjoint Wilson line in $SU(2)$ lattice gauge theory}
\author{Joe Kiskis}
\address{
Department of Physics and Institute of Theoretical Dynamics,\\
University of California, Davis, CA 95616, USA\\
jekucd@ucdhep.ucdavis.edu  }
\author{Pavlos Vranas}
\address{
SCRI, Florida State University, Tallahassee, FL 32306, USA   }
\date{\today}
\maketitle
\begin{abstract}
The behavior of the adjoint Wilson line in finite-temperature, $SU(2)$, lattice
gauge theory is discussed.  The expectation value of the line and the
associated excess free energy reveal the response of the finite-temperature
gauge field to the presence of an adjoint source. The value of the adjoint line
at the critical point of the deconfining phase transition is highlighted. This
is not calculable in weak or strong coupling. It receives contributions from
all scales and is nonanalytic at the critical point. We determine the
general form of the free energy. It includes a linearly divergent term that is
perturbative in the bare coupling and a finite, nonperturbative piece. We use a
simple flux tube model to estimate the value of the nonperturbative piece. This
provides the normalization needed to estimate the behavior of the line as one
moves along the critical curve into the weak coupling region.
\end{abstract}
\pacs{11.15.Ha, 11.15.Me, 11.15.Tk}
\narrowtext

\section{Introduction}

The context of this paper is the finite-temperature, deconfining phase
transition of pure-glue, $SU(2)$, lattice gauge theory. Many of the properties
of this transition are now well-established \cite{r0}. It is second order and
is in the universality class of the three-dimensional Ising model. The order
parameter is the expectation value of the Wilson line in the fundamental
representation. This is a measure of the response of the gauge system to the
presence of a nondynamical, static, external quark source. The response of the
system to the presence of sources in other representations has also been
studied \cite{r1,kis90,fmrsw} through the finite-temperature expectation
value of Wilson lines in higher dimensional representations.
The adjoint and $3/2$ have received most of the attention.

The expectation value of the Wilson line at temperature $T$ in representation
$J$ is the exponential of the corresponding excess free energy
\begin{equation}
  \langle L_J \rangle = e^{-F_J/T}.
\end{equation}
For the fundamental representation in the confined phase, the free energy is
infinite, and the expectation value is zero. In the deconfined phase, the free
energy is finite, and the expectation value is finite. It is believed that the
infinite free energy in the confined phase is due to a tube of color electric
flux with finite energy per unit length that extends from the source to
infinity.

Earlier work \cite{kis90,fmrsw} indicates
that the adjoint line is nonzero at all finite temperatures. Furthermore, if
$t$ is the deviation of the temperature or of gauge coupling from the critical
value, then the adjoint line has a
singular behavior at the critical point that is determined by the specific heat
exponent $\alpha$ \cite{kis90,fmrsw}.
\begin{equation}
  \langle L_1 \rangle = a+b_{\pm} |t|^{1-\alpha}.         \label{e1}
\end{equation}
In the confined phase, the adjoint source can be screened by a loop of flux of
finite length. Nevertheless, the nonanalytic behavior of
$\langle L_1 \rangle$ also reflects the diverging
fluctuations of the flux at the critical point.

{}From now on, we will be discussing the adjoint line and will drop the
subscript
``1''.

The purpose of this paper is to discuss the expectation value of the adjoint
line in more detail. It focuses on the magnitude of $\langle L \rangle$ and
especially upon its value at the critical point $\langle L \rangle _c$. The
expectation value is a function of two variables $N_T$ and $g^2$ or,
equivalently, $\Lambda$
and $T$. On the critical curve in either of these planes,
$\langle L \rangle _c$ is a function of a single variable which is
essentially the cutoff. As the discussion will show, $\langle L \rangle _c$
decreases rapidly as the cutoff is increased. This is an important limit on
attempts to verify (\ref{e1}) closer to the continuum limit. It will be
apparent that there is some interesting physics in producing the estimates
for the magnitude of $\langle L \rangle$. One reason is that, as a
matter of principle, neither weak nor strong coupling expansions can be used in
a straightforward way to calculate $\langle L \rangle_c$. The critical point
is just exactly the point at which the expansions must fail.
Furthermore, the values for $\langle L \rangle$ given by
the first terms in
either expansion near the actual critical point are far from the values
given by numerical simulation. For example, at $N_T=2$ and $4/g_{c}^2=1.87$,
the
Monte Carlo result is about $0.025$ while first order strong coupling
gives $0.009$, and first order weak coupling gives $-0.001$.
At $N_T=4$ and $4/g_{c}^2=2.296$, where the
Monte Carlo result is about $0.01$, first order weak coupling
gives $-0.63$.

To make progress, we will need a different approach. First, the general form of
$F$ will be established. With spatial cutoff $a$, coupling $g$ at scale $a$,
temperature $T$, and critical temperature $T_c$,
\begin{equation}
  F=\frac{1}{a} c(g^2)+ T_c d(T/T_c).          \label{e20}
\end{equation}
The first contribution is the linearly divergent self energy.
The function $c$ has a power series expansion. Its first term is the $O(g^2)$
Coulomb self energy. In a free gauge theory, that is all there would be to $F$.
Interactions give higher order corrections to the coefficient $c$. They also
give rise to the finite function $d$, which contains effects from the
nonperturbative scale $T_c \sim \Lambda_{QCD}$.

To estimate $d$, we will have to add some physical ideas and employ models
that are not systematic approximations. In one way of looking at it,
$d$ includes the binding of a gluon to the adjoint source to screen it.
Alternatively, one can think in terms of the flux tube model \cite{r2,knv2}
in which the phase transition is driven by the energy-entropy competition of
the flux tube. In such a picture, let $R$ be the return probability of the flux
tube, and let $A$ be the cost of attaching it to the line. Then we can write
\begin{equation}
  \langle L \rangle_c \sim AR.
\end{equation}
In the strong coupling region, $A$ and $R$ can be estimated.

Section II discusses the general form in (\ref{e20}), while Sec.\ III considers
the weak coupling expansion of $c(g^2)$. Section IV has discussion of the flux
tube model and its surprisingly good predictions for the critical coupling.
Finally, the main work is in Sec.\ V where the estimates of $d(0)$ and $d(1)$
are made. Section VI concludes the paper with a reiteration of the main
points, which can also be stated now:
Three methods with somewhat different approaches to the physics lead
to equivalent results:
\begin{equation}
  \langle L \rangle_c  =  e^{-[N_T c(g^2) +  d(1) ]}
\end{equation}
with
\begin{equation}
  c \approx (8/3) [ 0.351 / (4/g^2) + .294 / (4/g^2)^2 ]
\end{equation}
and
\begin{equation}
  e^{-d(1)} \approx 0.1  .
\end{equation}

\section{General form of $F$}

We will concentrate our attention on the free energy $F$ rather than on the
line. The reasons for this will become apparent as the discussion proceeds.

The first task is to establish the form
\begin{equation}
  F=\frac{1}{a} c(g^2)+ T_c d(T/T_c).        \label{e2}
\end{equation}
This can be written in a number of different ways depending upon whether one
prefers a notation suggested by a lattice or continuum cutoff and whether one
prefers to use only dimensionless variables.
In dimensionless lattice units, the form is
\begin{equation}
  aF = c(g^2) + d(N_{T_c}/N_T)/N_{T_c} \label{e11}.
\end{equation}

The form in (\ref{e2}) is essentially the same
as that derived by Dotsenko and Vergeles \cite{r3}. They treated
the case of a Wilson loop at zero temperature and showed that
\begin{equation}
  \langle W(C) \rangle = e^{\psi(C)}.
\end{equation}
$\psi$ contains a term that is linearly divergent and proportional to the
length
of the loop $C$. The coefficient of the linear divergence is a finite function
of the {\em bare} coupling $g$. It does not contain terms with logarithmic
divergence. After this term, which is proportional to the length of the loop,
is subtracted from $\psi$, the remainder is a finite function of the
{\em renormalized} coupling $g(\mu)$. We assume that the proof that they have
given at zero temperature applies at finite temperature also.
Then the finite term in $F$ has the form
\begin{equation}
  T d_1(T,g^2(\mu),\mu) = T d_2(T/\mu,g^2(\mu))  .
\end{equation}
When $\mu$ is chosen at $T_c$, it becomes
\begin{equation}
 T d_2(T/T_c,g^2(T_c)) = T_c d(T/T_c),
\end{equation}
since $g^2(T_c)$ is a fixed number of order one.
We have chosen to extract a $T_c$ in the final form rather than a $T$ because
it seems likely that this contribution is not zero at $T=0$, i.e. $d(0)$ is
finite. Our primary interest is in the critical value $d(1)$.

The linear divergence comes from the integration region where all of the points
at which the gluons attach to
the line come together, and the propagators have their $T=0$ short distance
structure. Thus, the coefficient of the linear divergence is independent of
$T$. If the renormalized charge is eliminated in favor of the scale $T_c$, then
$c$ must be a function of the dimensionless combination $aT_c$. However, this
ratio of the cutoff and physical scales is determined by the bare coupling
$g$ (the coupling at momentum scale $1/a$). Thus, $c$ is a finite function of
the bare coupling as indicated in (\ref{e2}).

For an indication of why $F$ is the preferred quantity, recall that
\begin{equation}
 \langle L \rangle = e^{-N_T a F} ,      \label{e13}
\end{equation}
that $c \sim g^2$, and that $g^2 \sim 1/\ln(N_T)$ on the weak coupling part of
the critical line. Thus, it is more convenient to deal with $F$ and then deduce
the behavior of $\langle L \rangle$.

At this point, we can see how our problem takes shape.
Along the critical line, $N_{T_c}$ increases as we move into weak coupling.
The function $c$ has a perturbative expansion and contributes a decreasing
factor to (\ref{e13}).
The function $d(N_{T_{c}}/N_{T})$ has contributions from short and long
distance. However, on the critical curve, we need $d(1)$, which is
inherently nonperturbative. It contributes a cutoff-independent factor
${\cal L}(1) \equiv e^{-d(1)}$ to $\langle L \rangle _c$. Thus, there are two
factors. The first is calculable and gives a rapid decrease of
$\langle L \rangle _c$ with decreasing lattice spacing and increasing $N_T$.
The second factor is nonperturbative and independent of the cutoff.

The first two terms in $c$ can be estimated from the the calculations of Heller
and Karsch \cite{r4}. To estimate $d(1)$, we will make use of simple models and
matching to the strong coupling region.

\section{Weak coupling and $\lowercase{c}(\lowercase{g}^2)$}

The calculations that Heller and Karsch \cite{r4} have done allow us to
estimate the first two terms in $c$. We use the data in their Table 5.
Although $c$ is an essentially short distance quantity, it should be noted
that there are some possible finite-volume problems. The finite-volume,
constant mode of the gauge field cannot be handled straightforwardly in
perturbation theory. It it has not been included in the calculations of
Ref.\ \cite{r4}. The volume dependence of the numbers that they
have computed gives a rough estimate for the effect of the deleted
mode and the finite-volume corrections. The spatial lattice sizes are
$8^3$, $10^3$, and $12^3$. The effect on the leading term in $c$ is at
the 2\% level. However, it can be seen from their table that there is a
substantial $N_S$ variation in the $O(g^4)$ numbers.
When they are used to compute the coefficient of $g^4$ in
$c$, the finite-size effects persist.

A closer analysis of the numbers at nearly fixed $TL=N_S/N_T$ shows that
there are also significant finite-spacing effects. At infinite cutoff, the
coefficient of the linear divergence is independent of $TL$. Thus, our best
estimate of the $O(g^4)$ contribution to the coefficient will
come from the largest $N_T$ at fixed $N_S/N_T$ but without regard for the
particular, fixed value of that ratio. The $10^3 \times 6$ and $12^3 \times 7$
numbers combine to give the ``$0.294$'' in (\ref{e4}) below.
This could easily be off by 20\%.
However, an error of this size will not have a
significant impact on the discussion to follow, in which other, larger
uncertainties arise.
This procedure gives our estimate
\begin{equation}
  c \approx (8/3) [ 0.351 / (4/g^2) + 0.294 / (4/g^2)^2 ]  .  \label{e4}
\end{equation}

The factor $8/3$ is needed to change from the fundamental to the
adjoint representation of $SU(2)$. In higher order, the relationship is not the
simple multiplicative factor that it is here.
The first term in $c$ is the lowest order contribution to the linearly
divergent Coulomb energy. It would be the whole answer if the theory were free.
The interactions correct the coefficient of the linear divergence through
interactions on the scale of the cutoff. This is the origin of the second term
in (\ref{e4}). However, the interactions also correct the energy on all other
scales. These are finite changes and are contained in the term $d$ in $F$. Most
of our attention will be focused on estimating these effects.

\section{Flux tube model for the phase transition}

Long ago Patel \cite{r2} proposed a flux tube model for the phase transition.
In this
picture, the phase transition is driven by the energy-entropy competition of
the
flux tube attached to a color source. This picture can be used to estimate the
position of the phase transition. We find that it does surprisingly well. This
suggests that the model is capturing some important physics and that it can be
used for other estimates in later sections.

Let the length of a basic unit of the flux
tube be $\tilde{a}$. In the strong coupling region, $\tilde{a}$ is the
lattice spacing $a$.
In weak coupling, it is of the order of the diameter of the flux tube, which in
turn is of the same order as the nonperturbative scale $1/\Lambda_{QCD}$
and the inverse root string tension $1/\surd(\sigma)$.
We will use $\tilde{a}=1/\surd(\sigma)$ in the weak coupling region.

Each unit of flux, has an associated energy $\sigma \tilde{a}$ and
Boltzmann weight $e^{-\sigma \tilde{a} /T}$.
The entropy factor for each unit is $\gamma$. For a free random walk
of flux in three
dimensions, there are six possible directions for each step, so $\gamma = 6$.
However, the flux tube has repulsive self interactions at crossings that
suggest the use of a self-avoiding walk. In that case, a simple guess for
$\gamma$ is $5$. Numerical simulations of self-avoiding walks give
$\gamma \approx 4.68$ \cite{r6}. We will use
that value. In other work \cite{knv2}, we have shown that the
fractal dimension of the flux tube at $T_c$ is near the dimension of a
self-avoiding walk.

The phase transition occurs when the temperature has increased
to the point that the product of the energy and entropy factors for each step
is equal to one.
\begin{equation}
 \gamma e^{-\sigma \tilde{a} /T_c} = 1
\end{equation}
In strong coupling, where $\sigma a^2 \approx \ln g^2$, that gives
\begin{equation}
   (g_{c}^2)^{N_T} = \gamma     \label{e8}
\end{equation}
or
\begin{equation}
   4/g_{c}^2 = 4 (1/\gamma)^{1/{N_T}} .   \label{e7}
\end{equation}
For $N_T=2$, this gives $4/g_{c}^2 = 1.85$. The Monte Carlo value is
$1.87$ \cite{mc1}.
For $N_T=1$, (\ref{e7}) and Monte Carlo give $0.855$ and $0.873$ \cite{mc2},
respectively.

In weak coupling a rearrangement of (\ref{e8}) gives,
\begin{equation}
   \ln\gamma = \sigma \tilde{a} /T_c = \surd(\sigma)/T_c  .
                                       \label{e9}
\end{equation}
A Monte Carlo calculation \cite{mc0} of the ratio on the right gave $1.45$.
On the other hand, the far left term is $1.54$. This is acceptable agreement
given the rough nature of the estimates.
For $N_T=3$, the phase transition is at $4/g^2=2.177$ \cite{mc0},  which is
just into  the weak coupling region. At this coupling, the square root of the
string tension is $0.492$ \cite{mc0}. This gives
$N_{T_c} \surd(\sigma a^2)=1.48$, which also compares reasonably with
$\ln \gamma = 1.54$ in (\ref{e9}).

The crossover between the strong and weak coupling regions is
around $g^2 \approx 4/g^2 \approx 2$. Thus, the $N_T=2$ phase transition is in
the strong coupling region but near the crossover.
At $4/g^2 = 1.87$, $\sigma a^2 = \ln g^2$ gives
$1/\surd(\sigma a^2) \approx 1.15$ so that the lattice spacing is only
slightly larger than the length $1/\surd(\sigma)$, which is the diameter of the
flux tube.
Thus, at this point, the lattice cutoff is not eliminating a
great many flux tube configurations.
A reasonable match of strong and weak coupling quantities
in this region seems possible.

One may worry that the strong coupling flux tube picture of the phase
transition that uses the lowest order result $\sigma a^2 = \ln g^2$
for the $T=0$ string tension
is too rough. However, the first
correction to the string tension comes from graphs with four extra plaquettes.
This is of relative order $(g^2)^{-4}$. Even at the crossover $g^2\approx 2$,
this is just $1/16$.

At this point, one may fairly inquire into the physical effects that have been
neglected in the flux tube picture. The model asserts that the important
finite temperature fluctuations of the gauge field are large scale
fluctuations in the
configuration of the flux tube that is attached to an external source. The flux
tube is treated as a structureless object that has no internal excitations when
the temperature is increased. As indicated, this may be a good approximation in
strong coupling. In weak coupling, where the flux tube has a finite,
nonperturbative size, this is a simplification that could be incorrect. In
addition, the picture does not explicitly include the thermal fluctuations that
are closed loops of flux or the more mundane excitations of short wavelength
modes. The latter may not be too important
since they are not highly excited at $T = T_c$.
The loops of flux that do not intersect the flux tube connected to the line
cancel in the ratio that is the {\em excess} free energy of the static source.
Part of the effect of those that do intersect may be already included in
using the self-avoiding walk value $\gamma = 4.68$.

\section{Estimates of $\lowercase{d}(0)$ and $\lowercase{d}(1)$ }

\subsection{$d(0)$ and the glueball mass}

One rough estimate of $d(0)$ comes from the glueball mass.
Let us view the glueball as a bound state of two gluons, each of which carries
the adjoint representation of the gauge group. Then the glueball mass is an
indication of the energy associated with the response of the vacuum to an
adjoint source. The problem, of course, is that the gluons are ``massless''
while the line is infinitely massive. For nonrelativistic bound states,
the system
with one infinitely massive constituent has twice the binding energy and one
half the radius of the equal mass system.
On the other hand, a simple bag model calculation that includes only volume and
kinetic energies gives a glueball mass a
factor of $2^{3/4}$ larger than the system with one infinitely massive
constituent \cite{bag}. If the effects of
vacuum fluctuations and attractive color hyperfine interactions were included,
they would act oppositely so that the ratio would not change greatly.
This suggests that $d(0)$
might be within a factor of two of the ratio of the glueball mass to the
critical temperature. The latter has been measured at $5.6$ \cite{mc0}. Thus,
for a rough guess with large uncertainty, put
\begin{equation}
 d(0) \approx 5.6 .                  \label{e5}
\end{equation}

\subsection{Strong coupling estimate of $d(0)$}

It is easy enough to get the lowest order strong coupling contribution to
$aF(0)$. It comes from the strong coupling diagram that is a
$1 \times 1 \times 1$ tube of plaquettes in the inverse temperature direction:
\begin{equation}
 aF(0) = 4 \ln g^2.
\end{equation}
In units of $T_c$, this is
\begin{equation}
 F(0)/T_c = 4 N_{T_c} \ln g^2.
\end{equation}
On the other hand, we know that
\begin{equation}
 (g^2)^{N_{T_c}}=\gamma
\end{equation}
so
\begin{equation}
 F(0)/T_c = 4 \ln \gamma \approx 6.2.
\end{equation}
At this level of approximation, the result is independent of $g^2$. Notice that
it is similar to the number in (\ref{e5}). In the strong coupling region, there
is nothing that corresponds to the short distance effects in $c$.
Thus, the strong coupling
$F(0)/T_c$ can also be thought of as $d(0)$.

\subsection{$d(1)$ from simple matching}

Recall that we are looking for an estimate of
$F(T_c)$ that is useful for decreasing $g^2$ and increasing $N_T$.
This means that we need the
value of $d(1)$ in $aF(T_c)= c(g^2) + d(1)/N_{T_c}$.

\subsubsection{Monte Carlo}

The simplest thing to do is to determine $F(T_c)$ from a
Monte Carlo measurement of $\langle L \rangle_c$ at one point in the weak
coupling region where $c$ can be reliably computed. With $F$ and $c$ given, $d$
is determined. Unfortunately, there are no numerical experiments that have
carefully looked at the critical value of $\langle L \rangle$ with small
coupling and large $N_T$. It is easy to see from the published Monte Carlo
data or from our discussion that $\langle L \rangle_c$ will
be small and somewhat difficult to measure at large $N_T$.
At both $N_T=1$ and $2$, $\langle L \rangle_c \approx 0.025$ \cite{kis90,vra}.
The values at $N_T=4$ and $6$ are smaller and more uncertain---around
$0.01$ at $N_T=4$ \cite{rs} and $0.001$ to $0.0014$ at $N_T=6$
\cite{fmrsw}.
If, for the sake of discussion, we use these $N_T=4$ and $6$ numbers, then the
corresponding values for
$d(1)$ are $2.38$ and $3.79$ to $3.45$.
The values for
${\cal L}(1)$ are $0.093$ and $0.022$ to $0.031$.

\subsubsection{Strong coupling form}

The next simplest thing to do is to match the weak coupling form (\ref{e11})
to a strong
coupling form at the crossover around $g^2=2$.
As indicated in the introduction, it is not possible to calculate at $T_c$
in a straightforward expansion. Instead, we will employ a very
simple model. It is inspired by the flux tube picture of the transition. We
assume that the most important contribution at $T_c$ is captured in  a very
limited subset of the strong coupling diagrams. These diagrams are certain
generalized cylinders of plaquettes in the fundamental  representation. They
are invariant under translations along the  inverse temperature axis. On each
three-dimensional slice, they consist of a self-avoiding loop that begins and
ends at the location of the adjoint source. This model goes a bit too far in
that the flux tube has repulsive self interactions but is not strictly
self-avoiding. However, in a related context, we have found \cite{knv2}
that the fractal
dimension of any field theory  walk with repulsive interactions and
interactions with background loops is close to the dimension $1.69$ of a
strictly self-avoiding walk. So perhaps some of the effects that have been
neglected are partly accounted for in the assumption of a self-avoiding
configuration of flux.

With this restriction on the allowed graphs, it is possible to sum up their
contribution at $T_c$. We did this in the following way: The asymptotic form
for the number of closed, self-avoiding configurations is \cite{r6}
\begin{equation}
 N(l) \sim l^{-7/4} \gamma^l .                    \label{e10}
\end{equation}
In addition, each configuration is weighted by the cost of a strip of $N_T$
plaquettes raised to a power that is the length of the path
\begin{equation}
  [(1/g^2)^{N_T}]^l .
\end{equation}
It follows that the critical point is at
\begin{equation}
  (1/g^2)^{N_T} = 1/\gamma
\end{equation}
as indicated previously.

The sum to be done is
\begin{equation}
 R = \sum_{l=4}^{\infty} N(l) (1/\gamma)^l  .
\end{equation}
For the terms through $l=12$, we used the explicit numbers for $N(l)$
given in \cite{fs} and, for the rest, the asymptotic form (\ref{e10}).
The multiplicative constant implicit in (\ref{e10}) was adjusted to match the
known value at $l=12$. This gives $R=0.078$.
To get the value
for $\langle L \rangle_c$, we must divide this by the ``3''
that relates the Wilson line to the adjoint character. The result is
\begin{equation}
   \langle L \rangle_c \approx R/3 \approx 0.026  .
\end{equation}
Notice that this agrees with the $N_T=1$ and $2$ Monte Carlo results.

\subsubsection{Matching}

Now select $d(1)$ so that as $g_{c}^2$ approaches the
crossover region from the weak coupling side, then $\langle L \rangle_c$
approaches $0.026$. The equations (\ref{e11}) and (\ref{e4}) are used.
Using $N_T=2$, $4/g^2=1.87$, and
$\langle L \rangle_c=0.026$, gives $d(1)=2.20$.

The sensitivity of $d(1)$ to the details of the matching will be discussed in
Sec.\ V.F. For now, we will accept
\begin{equation}
  d(1) \approx 2.2     \label{e31}
\end{equation}
as a reasonable estimate with considerable uncertainty.
This can be compared with the estimate $d(0)=5.6$ that came from the
glueball mass. As one would guess, it appears that $d(0)$ and $d(1)$ are of
the same order of magnitude.

\subsection{Return probability approach to $d(1)$}

This is an alternative approach to the flux tube picture of the phase
transition.
It eventually leads to the same calculation as did the previous method.
There is a loop of flux
attached to the adjoint source. As the critical point is approached, this flux
has increasing fluctuations. The number $0.026$ computed above can be thought
of
as one third of the sum over lengths of the return probabilities of
self-avoiding configurations of the flux tube.

In Sec.\ V.C.2, we discussed this in the strong coupling region.
To use the same picture in weak coupling, an
elaboration is required. There is a cost associated with attaching the flux
tube to the point source. The diameter of the flux tube is of the order of the
inverse root of the string tension, which is many lattice spacings in the weak
coupling region. The source is a point on the lattice. This mismatch of scales
has an associated cost that we will call $A$.
With this in mind, we consider the form
\begin{equation}
 \langle L \rangle = \frac{1}{3} A R .     \label{e12}
\end{equation}
Again, the $1/3$ is the conventional factor in the relation between the line
and the adjoint character.

In the strong coupling region,
the diameter of the flux tube is smaller than a lattice spacing so that
$A=1$, explicitly, in the lowest order strong coupling model described above.
The first correction to $A=1$ is $O(N_T(g^2)^{-6})$ (when the line
excludes two touching boxes on the sheet), which is small for the strong
coupling values of $N_T$ and $g^2$ that are relevant.
$A$ is unknown in weak coupling
because it contains contributions from scales from the lattice spacing on up
to the the size of the flux tube. However, as the crossover is approached these
scales converge and $A \rightarrow  1$.

Now consider the return probability $R$ in the weak coupling region at the
phase transition. Loops of arbitrarily large size contribute to $R$. This is
large distance physics. We assume that strong coupling gives an adequate
description of large distances. Thus, we use $R=0.078$ in both regions.

To determine $d(1)$, use Eqs. (\ref{e11}), (\ref{e13}), and (\ref{e4})
on the left hand side of (\ref{e12}) and evaluate it
in the crossover region with $A \approx 1$. Evidently,
this leads to the previous result. However, the discussion has provided a bit
more insight into the nature of the physical approximations that are involved.

\subsection{ $d(1)$ from matching $F(0)$ }

This is another way to look at the physics of the problem.
We assume that for {\em all} couplings, the change in the free energy as the
temperature is increased from zero to $T_c$ is mainly due to the increasing
large scale fluctuations of the flux tube and further that these large scales
are adequately described by our model of the {\em strong} coupling flux tube.
This will give us $d(1)-d(0)$. To get $d(0)$, we require continuity of the zero
temperature free energy $F(0)$ through the crossover region.
Although this
approach is significantly different from the two already discussed, it will
turn out to be nearly equivalent in its net result.

In the first step, we adjust $d(0)$ so that the weak coupling expression for
$F(0)$ approaches the strong coupling value as $g^2$ is increased.
Specifically, we require
\begin{equation}
  c/a + T_c d(0) = (4/a) \ln g^2
\end{equation}
or
\begin{equation}
  d(0) = 4 N_{T_c} \ln g^2 - N_{T_c} c(g^2)    \label{e14}
\end{equation}
at the crossover. For $g^2=2$ and $N_T=2.3$ \cite{r7}, this gives $d(0)=4.85$.
It can be compared directly with the value $d(0) \approx 5.6$ obtained from
the glueball mass. Given the uncertainty in the latter, these are not
inconsistent.

The second step asserts that the change in the free energy
\begin{equation}
 [ F(T_c)-F(0) ]/T_c = d(1) - d(0)
\end{equation}
in {\em weak} coupling can be approximated by the {\em strong} coupling flux
tube value. Our model for the flux tube gives a value for the change in the
free energy that becomes independent of $g^2$ on the strong
coupling side of the crossover.
\begin{eqnarray}
 [ F(T_c)-F(0) ]/T_c & = & -\ln[\langle L \rangle_c ]_{sc} -
                                                  4 N_{T_c} \ln g^2 \\
           & = & -\ln[\langle L \rangle_c ]_{sc} -4 \ln \gamma \\
           & = & -\ln(0.026) - 4 \ln(4.68) \\
           & = & -2.52
                                             \label{e33}
\end{eqnarray}
and
\begin{equation}
   d(1) = d(0) + [d(1) - d(0)] \approx 4.85 - 2.52 \approx 2.3  .
\end{equation}
This is not quite the same as the value $2.2$ in (\ref{e31}). The method of the
present section would give precisely the result of (\ref{e31}) if the values
for $4 N_{T_c} \ln g^2$ in  (\ref{e14}) and (\ref{e33}) had been the same.
In (\ref{e14}), we worked at the crossover for the best estimate of $d(0)$.
However, in the spirit of the method, a value well into the strong coupling
region should be used in (\ref{e33}). So as a matter of principle, this method
is not the same as that of Sec.\ V.C. However, $4 N_{T_c} \ln g^2$ is a slowly
varying function over the range from the crossover and into strong coupling.
Thus, in practice the values for $d(1)$ end up rather close, The difference
is not significant considering the rough nature of our approximations.

\subsection{Sensitivity to the matching point}

There is some freedom in the details of the matching in the crossover region.
We will consider these for the procedure of Sec.\ V.C.3. Similar considerations
apply to the other methods. Since all of the methods are equivalent or nearly
equivalent to that of Sec.\ V.C.3, it is sufficient to discuss that case.

In Table \ref{table1}, we have chosen representative values for $N_T$,
$4/g_{c}^2$, and  $\langle L \rangle_c$.
For the noninteger values of $N_T$, $4/g_{c}^2$ was calculated from (\ref{e7}).
The results for $d(1)$ and ${\cal L}(1)$ follow from
(\ref{e11}), (\ref{e13}), and (\ref{e4}).  We conclude that ${\cal L}(1)$ is
somewhat sensitive to the details of the matching. However, except for the
$N_T=3$ cases, for which the $\langle L \rangle_c$ values were arbitrarily
chosen, the results for $d(1)$ and ${\cal L}(1)$ are around $2.1$ to $2.5$
and $0.09$ to $0.12$, respectively.

\section{Net result}

While each of the methods end with essentially the same result, the physics
is viewed
from a different perspective in each of the cases. This helps us to understand
the true nature of the approximation.

The final expression is
\begin{eqnarray}
  \langle L \rangle & = & e^{-[N_T c(g^2) + (T_c/T) d(T/T_c) ] }\\
                    & = & e^{-N_T c(g^2)} {\cal L} (T/T_c) .
\end{eqnarray}
The function $c$ is given in (\ref{e4}). At the critical point,
\begin{equation}
  \langle L \rangle_c = e^{-N_T c(g^2)} {\cal L} (1)
\end{equation}
with
\begin{equation}
  {\cal L} (1) \approx  0.09 \text{ to } 0.12  .
\end{equation}

This is several times higher than the $N_T=6$ Monte Carlo numbers that were
mentioned in Sec.\ V.C.1. However, we feel that those numbers must be
considered
tentative until confirmed by a simulation that is specifically focused on
$\langle L \rangle_c$. The main technical weakness of our work is the
expression for $c(g^2)$. To reduce the uncertainties, it would be useful
to have an infinite volume, $O(g^6)$ calculation of $c$.

\acknowledgments

P.V. would like to thank U. Heller for useful discussions.
This research was supported by the United States Department of Energy.
For P.V., the grant numbers are DE-FG05-85ER250000 and DE-FG05-92ER40724.

\begin{table}
\caption{Matching points.}
\label{table1}
\begin{tabular}{lllll}
 $N_T$ & $4/g^2$ & $\langle L_1 \rangle$ & $d(1)$ & ${\cal L}(1)$\\ \tableline
   2.0 &   1.87  &         0.026         &  2.20  &   0.11       \\
   2.2 &   1.98  &         0.026         &  2.17  &   0.11       \\
   2.3 &   2.0   &         0.026         &  2.12  &   0.12       \\
   2.4 &   2.10  &         0.026         &  2.15  &   0.12       \\
   3.0 &   2.177 &         0.026         &  1.86  &   0.16       \\
   2.0 &   1.87  &         0.020         &  2.46  &   0.085      \\
   2.2 &   1.98  &         0.020         &  2.43  &   0.088      \\
   2.3 &   2.0   &         0.020         &  2.38  &   0.092      \\
   2.4 &   2.10  &         0.020         &  2.42  &   0.089      \\
   3.0 &   2.177 &         0.020         &  2.13  &   0.12       \\
   3.0 &   2.177 &         0.010         &  2.82  &   0.060      \\
\end{tabular}
\end{table}


\begin{references}
\bibitem{r7} The value $N_T=2.3$ for $g^2=2$ was obtained by interpolating
     between the
     Monte Carlo results at integer values of $N_T$. Eq.\ (\ref{e8}) gives
     essentially the same result.
\end{references}
\end{document}